\def\par{}
\newcommand\be{\begin{equation}}
\newcommand\ee{\end{equation}}
\def\bea{\begin{eqnarray}}
\def\eea{\end{eqnarray}}
\def\N{{\frak L}}
\def\A{\frak A}
\def\bA{\overline{{\frak A}}}
\def\Bpm{{{\frak A}^\pm_\N}}
\def\bBpm{{\overline{{\frak A}}^\pm_\N}}
\def\B{{\frak A}^+_\N}
\def\bB{\overline{{\frak A}}^+_\N}
\def\tB{{\frak A}^-_\N}
\def\tbB{\overline{{\frak A}}^-_\N}
\def\a{\frak a}
\def\ba{\overline{{\frak a}}}
\def\b{{\frak a}^+_\N}
\def\bb{\overline{{\frak a}}^+_\N}
\def\tb{{\frak a}^-_\N}

\def\e{{\rm e}}
\def\ps{\psi_\N(x)}
\def\bps{\overline{\psi}_\N(x)}

\def\Annp{Ann. Physik\ }

\def\jpj{J. Phys. Soc. Japan\ }

\def\ptp{Prog. Theor. Phys.\ }

\def\jsp{J. Stat. Phys.\ }

\def\npb{Nucl. Phys. B\ }

\def\pla{Phys. Lett. A\ }

\def\prb{Phys. Rev. B\ }

\def\prl{Phys. Rev. Lett.\ }

\def\zpb{Z. Phys. B\ }
\def\zp{Z. Phys.\ }
\documentclass[11pt,a4paper,twoside]{article}
\usepackage[utf8]{inputenc}
\usepackage{cite}
\usepackage{graphicx}
\usepackage{amsmath,amsfonts,amssymb}
\usepackage{epsfig}

\title{The complete conformal spectrum of a $sl(2|1)$ invariant network model
and logarithmic corrections}

\author{Britta Aufgebauer, Michael Brockmann, Win Nuding, Andreas Kl\"umper\\
 \parbox{0.9 \linewidth}{\vspace{0.4 \baselineskip}\centering
  Fachbereich C -- Physik, Bergische Universit\"at Wuppertal, \\ 42097
    Wuppertal, Germany}\\
\\
and\\
\\
Ara Sedrakyan\\
 \parbox{0.9 \linewidth}{\vspace{0.4 \baselineskip}\centering
{Yerevan Physics Institute}, {Theoretical Department},\\
{Br. Alikhanyan street 2}, {Yerevan 36, Armenia}}}
\date{\today}
\begin{document}
\maketitle
\begin{abstract}
We investigate the low temperature asymptotics and the finite size spectrum of
a class of Temperley-Lieb models. As reference system we use the spin-1/2
Heisenberg chain with anisotropy parameter $\Delta$ and twisted boundary
conditions. Special emphasis is placed on the study of logarithmic corrections
appearing in the case of $\Delta=1/2$ in the bulk susceptibility data and in
the low-energy spectrum yielding the conformal dimensions. For the $sl(2|1)$
invariant 3-state representation of the Temperley-Lieb algebra with
$\Delta=1/2$ we give the complete set of scaling dimensions which show huge
degeneracies.
\end{abstract}

\clearpage
\section{Introduction}
\label{sec:intro}
Many different, and at first sight unrelated physical phenomena like the
thermodynamics of quantum spin chains, critical properties of two-dimensional
classical vertex models and of supersymmetric network models are related on
fundamental mathematical grounds. Often, such relations are established by
mappings of quantum systems to classical systems, where the related
objects are operators like the Hamiltonian and the transfer matrix `living' in
the same Hilbert space of physical states.  Sometimes, the relations are less
obvious as the related objects are defined in rather different spaces, but
turn out to be equivalent representations of some underlying algebraic
structure. The structure relevant to our investigations is the so-called
Temperley-Lieb algebra \cite{TemperleyLieb71,Baxt82b,PMBook91}.

The seminal model for correlated quantum many-body systems and the standard
reference model of the Temperley-Lieb algebra is the well-known Heisenberg
model. In one spatial dimension the system with nearest-neighbour exchange is
exactly solvable for the spin-1/2 case \cite{Bethe31}.  Despite the
integrability a comprehensive theoretical treatment is still missing. A
notorious problem is posed by the calculation of correlation functions even at
zero temperature, not to mention the corresponding properties at finite
temperature. However, the asymptotics of the correlations in particular in the
ground-state and at low temperatures are quite well understood by a combination
of Bethe ansatz \cite{Baxt82b} and conformal field
theory \cite{BelaPZ84,Card88}.  At exactly zero temperature the spin-1/2
Heisenberg chain shows algebraic decay in its correlation functions and
thereby constitutes a quantum critical system.

Recently, the special point $\Delta=1/2$ of the anisotropic version of the
Heisenberg chain, the so-called spin-1/2 $XXZ$ chain, has attracted strong
interest as certain mathematical simplifications occur in the construction of
the eigenstates \cite{RazumovStrog01,RazumovStrog04,RazumovStrog05}. 
Also its classical counterpart, the six-vertex model with
suitable boundary conditions, allowed for new insight into many combinatorial
problems like for instance the alternating sign matrices and boxed plane
partitions, see e.g. \cite{BatchGN01,Kuperberg96} and references therein.

In the case $\Delta=1/2$ like for other special points, the
so-called roots of unity of the $XXZ$ chain, a crossover of critical exponents
leads to logarithmic corrections to the dominant critical
behaviour. A notable case is the isotropic model $\Delta=1$ with logarithmic
corrections to the low-temperature asymptotics, in particular for the
susceptibility\cite{EggertAT94}. For the case $\Delta=1/2$ strong logarithmic
corrections to the boundary susceptibility were found in \cite{SirkerBortz06}.

In this paper we report on a low-temperature analysis of the $\Delta=1/2$ bulk
properties that reproduce and confirm some of the findings
in \cite{Lukyanov1997,SirkerBortz06}. The main motivation of this analysis is
the applicability to the study of all low-lying excitations of the $XXZ$ chain
with arbitrary twisted boundary condition
allowing to calculate the energy levels beyond the conformal field theoretical
results.

The feasibility of such calculations is interesting because of applications to
statistical mechanical models of the spin quantum Hall
transition \cite{ChalkerCodd,Gade,Gruzberg}. The
localization-delocalization exponent was calculated in 
\cite{Gruzberg} from the mapping onto the two-dimensional percolation problem. 
The underlying system is a supersymmetric realization of the Temperley-Lieb
algebra. Another model for describing the spin quantum Hall
transition was introduced in \cite{Gade} and investigated extensively in
\cite{EssFrahmSaleur}, however, still leaving open important questions. One of
those questions concerns the relation of the two models.

For the network model \cite{Gruzberg} the complete classification of the
excited states resp.~the scaling dimensions was performed by a Coulomb gas
analysis in \cite{ReadSaleur01}. A combinatorial analysis based on a
decomposition of the partition function for the Potts model on finite tori in
terms of generalised characters can be found in \cite{RichardJacobsen}.

Here we present a systematic representation theoretical analysis of the
eigenvalues of the Temperley-Lieb Hamiltonian corresponding to the
network model on finite chains with periodic and $\pi$-twisted boundary
conditions. Our analysis is based on \cite{AufgebauerK10} where a
Temperley-Lieb equivalence with particular emphasis on the boundary conditions
was established for vertex-models. 
The precise mapping of the generic irreducible sub-representations of Temperley-Lieb
Hamiltonians with periodic boundary conditions to the Heisenberg chain with
suitable twisted boundary conditions allows for the determination of critical
properties. Also in \cite{AufgebauerK10}, the multiplicities of the
sub-representations were worked out. The gained knowledge of the complete
spectrum allowed for the computation of the thermodynamics of the
Hamiltonians.

In Section 2 of this paper an approach to the thermodynamics of the Heisenberg
chain on the basis of non-linear integral equations for just two auxiliary
functions is analysed \cite{KlumTH} (see also the related approaches by
\cite{Koma,Tak91,SuzukiAW90}). For $\Delta=1/2$ an analytical 
calculation of the leading universal terms to the free energy and the magnetic
susceptibility as well as the next-leading logarithmic corrections is
presented. In Section 3 we study the critical exponents of correlation
functions at zero temperature.  The leading conformal finite-size spectrum of
the Heisenberg chain with twisted boundary conditions and the next-leading
correction terms in the case $\Delta=1/2$ beyond the CFT results are derived.
In Section 4 we address the supersymmetric network model and calculate the
complete set of scaling dimensions as well as their multiplicities and
logarithmic corrections.

\section{Thermodynamics and logarithmic corrections}

We investigate the thermodynamical properties of the antiferromagnetic
isotropic spin-1/2 Heisenberg chain 
\begin{equation}
H = 2J\sum_{j=1}^L [S_j^xS_{j+1}^x+S_j^yS_{j+1}^y+\Delta S_j^zS_{j+1}^z]
\end{equation}
with anisotropy parameter $\Delta=\cos\gamma$ where $\gamma(\Delta=1/2)=\pi/3$.

\begin{figure}[!]
  \begin{center}
    \includegraphics[width=0.7\textwidth]{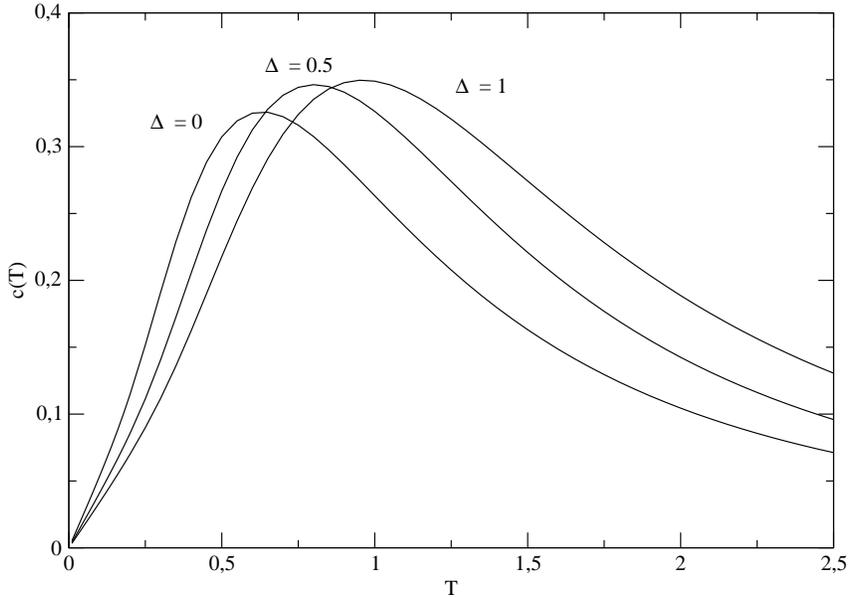}
    \caption{Specific heat $c(T)$ data versus temperature $T$ ($J=1$) for the
      spin-1/2 $XXZ$ chain with anisotropy $\Delta= 0, 0.5, 1$ and
    zero field $h=0$.}
    \label{fig:spec}
  \end{center}
\end{figure}

\begin{figure}[!htb]
  \begin{center}
    \includegraphics[width=0.7\textwidth]{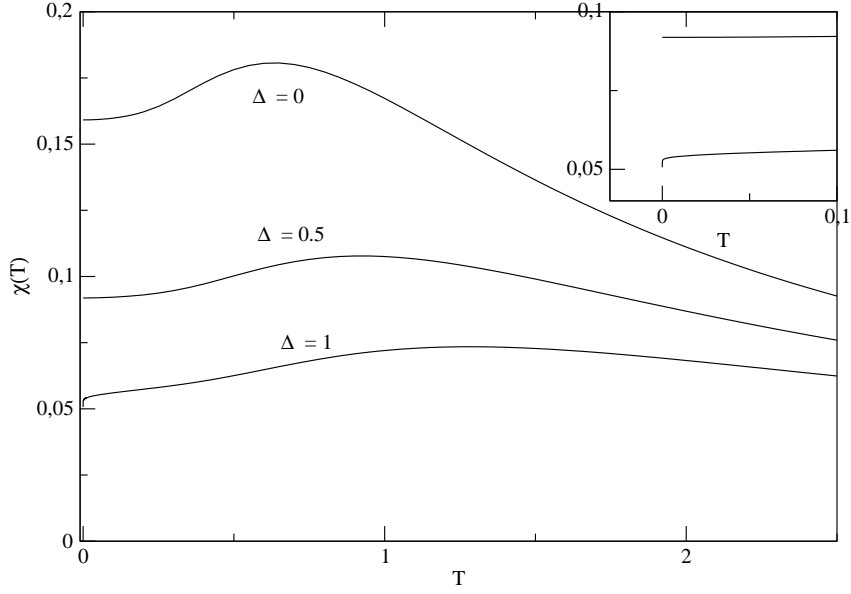}
    \caption{Susceptibility $\chi(T)$ data versus temperature $T$ ($J=1$) for
    the spin-1/2 $XXZ$ chain with anisotropy $\Delta= 0, 0.5, 1$ and
    zero field $h=0$. In the inset the low temperature behaviour of $\chi(T)$
    for $\Delta= 0.5$ and $1$ is shown.}
    \label{fig:susc}
  \end{center}
\end{figure}

We note that the elementary excitations (`spinons') have quasi-linear
energy-momentum dispersion with velocity
\begin{equation}
v=J \pi \frac{\sin\gamma}\gamma\to_{\Delta=1/2}\frac{3\sqrt{3}}2 J.
\end{equation}
The free energy per lattice site of the system is given by the following
set of non-linear integral equations \cite{KlumTH} for auxiliary functions
$\a$, and $\A=1+\a$
\begin{equation}
\log \frak a(x) 
= -\frac{v\beta}{\cosh x}
+\phi+\int_{-\infty}^\infty\left[k(x-y)\log \frak A(y)
-k(x-y-i\pi+i\epsilon)\log\overline {\frak A}(y)\right]{\rm d}y,
\label{IntEqxxx}
\end{equation}
and the corresponding equation for $\ba$, and $\bA=1+\ba$ are related to
(\ref{IntEqxxx}) by `complex conjugation', in particular by exchanging
$\a \leftrightarrow\ba$, and by changing the sign of $\phi\leftrightarrow
-\phi$ 
though being {\it real} as given by the external magnetic field
$\phi=\frac{\pi\beta h}{2(\pi-\gamma)}\to_{\Delta=1/2} \frac 34 \beta h$.
The integration kernel is defined by the Fourier integral
\begin{equation}
k(x)=\frac{1}{2\pi}\int_{-\infty}^{\infty}
\frac{\sinh\left(\frac{\pi}{\gamma}-2\right)\frac\pi 2 k}
{2\cosh\frac{\pi}{2}k\sinh\left(\frac{\pi}{\gamma}-1\right)\frac\pi 2 k}
\cos(kx)dk\to_{\Delta=1/2}\frac 1{2\pi^2}\frac x{\sinh x}.
\label{kernel}
\end{equation}
In terms of the solution $\A$ and $\bA$ to the integral equations the
free energy is given by
\begin{equation}
\beta f=\beta e_0-\frac{1}{ 2\pi}
\int_{-\infty}^\infty\frac{\log[\A\bA(x)]}{\cosh x}dx.
\label{eigenxxx}
\end{equation}
These equations are readily solved numerically for arbitrary fields and
temperatures by utilizing the difference type of the integral kernel and Fast
Fourier Transform.  In Figs.~\ref{fig:spec} and \ref{fig:susc} the results for
the specific heat and susceptibility are shown for zero external magnetic
field.  At low temperatures the specific heat shows linear asymptotics
$c(T)\simeq\frac{\pi c}{3v}T$ with velocity of the elementary excitations $v$
and central charge $c=1$. The zero temperature limit of the susceptibility is
$\chi_0=1/(2v(\pi-\gamma))$.

For $\Delta=1$ there are logarithmic corrections to the susceptibility
$\chi(T)$ at low temperatures as noticed in \cite{EggertAT94,OkaNom92} and to
the specific heat $c(T)$. For a detailed analysis of the isotropic point we
refer the reader to \cite{EggertAT94,Lukyanov1997,KlumperJohnston00}. We like
to point out that already for moderately low temperatures the correction terms
are noticeable for $\chi(T)$, see Fig.~\ref{fig:susc}.

There are other cases of the anisotropy parameter $\Delta$ with logarithmic
corrections at low temperatures. One of these cases is $\Delta=1/2$ as noticed
and analysed in \cite{Lukyanov1997,SirkerBortz06}. Already a glance to
Fig.~\ref{fig:susc} suggests that these correction terms must be of higher,
i.e.~less leading, order.  Here we like to present a completely analytical
low-temperature analysis (for $h<<T$) for the case $\Delta=1/2$ from the
non-linear integral equations.

With view to the asymptotical properties of the driving term on the RHS of
(\ref{IntEqxxx}) it is convenient to introduce the following scaling functions
\bea
\b(x)&=&\a(x+\N),\qquad \N:=\ln(v\beta),\nonumber\\
\tb(x)&=&\a(-x-\N),
\eea
approaching well-defined non-trivial limiting functions in the low-temperature
limit which satisfy
\bea
\log \b(x) 
&=& -2\e^{-x}+O\left({1}/{\beta^2}\right)
+\phi+\ps\nonumber\\
&&+\int_{-\N}^\infty\left[k(x-y)\log \B(y)
-k(x-y-i\pi+i\epsilon)\log\bB(y)\right]{\rm d}y, 
\label{IntEqxxxl}
\eea
where $\ps$ accounts for the contribution of the functions $\A$ and $\bA$
on the negative real axis
\begin{equation}
\ps=\int_{-\N}^\infty\left[k(x+y+2\N)\log \tB(y)
-k(x+y+2\N-i\pi+i\epsilon)\log\tbB(y)\right]{\rm d}y.
\label{ps}
\end{equation}
Using these functions the free energy at low temperatures
takes the form
\begin{equation}
f=e_0-\frac{1}{\pi v\beta^2}
\int_{-\N}^\infty\e^{-x}\log\left[\B\bB\tB\tbB(x)\right]dx
+O\left(1/\beta^4\right).
\label{finiteT}
\end{equation}
In order to analyse the low-temperature asymptotics in detail we
perform the following manipulation 
\bea
&&\int_{-\N}^\infty\left[(\log\b)'(\log\B)-(\log\b)(\log\B)'\right]dx
+\hbox{`conj'},
\nonumber\\
&&=\int_{-\N}^\infty\left[(-2\e^{-x}+\phi+\ps)'(\log\B)-
(-2\e^{-x}+\phi+\ps)(\log\B)'\right]dx+\hbox{`conj'},\nonumber\\
&&=2\int_{-\N}^\infty(2\e^{-x}+\ps')(\log\B)dx-
\phi\log\left(1+\e^{\frac 43\phi}\right)+\hbox{`conj'},
\label{eq1}
\eea
where in the second line we have inserted (\ref{IntEqxxxl}) and used the
symmetry of the integration kernel ($k_{ij}(x)=k_{ji}(-x)$ where indices 1 and
2 refer to functions $\a$, $\A$ and $\ba$, $\bA$ respectively). In the
third line we have performed an integration by parts with an explicit
evaluation of the contribution by the integral terminals using the limits
\bea
\log\b(\infty)&=&{+\frac 43\phi},\qquad \log\B(\infty)=
\log\left(1+\e^{+\frac 43\phi}\right),\nonumber\\
\log\bb(\infty)&=&{-\frac 43\phi},\qquad  
\log\bB(\infty)=\log\left(1+\e^{-\frac 43\phi}\right).
\label{limits}
\eea
Next, we show that the LHS of (\ref{eq1}) can be evaluated explicitly
as it is a definite integral of dilogarithmic type with known terminals
\bea
LHS&=&2\int_{-\N}^\infty(\log\b)'\log\B dx-
\log\b \log\B\Big|_{-\N}^\infty+\hbox{`conj'},\nonumber\\
&=&2\int_{-\infty}^{4\phi/3}\log(1+\e^z)dz-\frac 43\phi\log(1+\e^{\frac 43\phi})
+2\int_{-\infty}^{-4\phi/3}\log(1+\e^z)dz+\frac 43\phi\log(1+\e^{-\frac 43\phi})
,\nonumber\\
&=&\frac{\pi^2}{3},
\label{eq2}
\eea
where the second line is independent of $\phi$ (to be checked by 
differentiation) and the particular choice $\phi=0$ `directly' leads to 
the last line.

From the general expression (\ref{eigenxxx}) of the free energy $f$ in terms
of the functions $\A$ and $\bA$ the detailed functional dependence of $\A$ and
$\bA$ on the argument matters even for the calculation of the asymptotic
behaviour of $f$ at low temperature.  The usefulness of (\ref{eq1}) lies in
the fact that the integral of interest is related to another integral where
only the asymptotic behaviour of $\A$ and $\bA$ at large spectral parameter
enters.

Combining (\ref{eq1}) and (\ref{eq2}) we arrive at a formula particularly
suited for studying the asymptotic behaviour of (\ref{finiteT})
\begin{equation}
4\int_{-\N}^\infty\e^{-x}\log[\B\bB]dx=
\frac{\pi^2}{3}+\frac 43 \phi^2-2\int_{-\N}^\infty\left[\ps'\log\B+\bps'\log\bB
\right]dx.
\label{eq3}
\end{equation}
The first term on the right hand side determines the leading $O(T^2)$ term of
the free energy, see (\ref{finiteT}), the remaining integral is a higher order
term which can be calculated by explicit use of special properties of the
integral kernel (\ref{kernel}), especially the asymptotic behaviour
\begin{equation}
k(x)\simeq \frac 1{\pi^2}x\, \e^{-x}.
\end{equation}
This allows to approximate (\ref{ps}) by
\begin{equation}
\ps\simeq\bps\simeq\int_{-\N}^\infty k(x+y+2\N)\log \left[\tB(y)\tbB(y)\right]{\rm d}y,
\label{psasymp}
\end{equation}
leading to a symmetric double integral
\bea
&&\int_{-\N}^\infty\left[\ps'\log\B(x)+\bps'\log\bB(x)\right]dx\nonumber\\
&&\simeq\int_{-\N}^\infty\int_{-\N}^\infty
\log \left[\B(x)\bB(x)\right]k'(x+y+2\N)\log \left[\tB(y)\tbB(y)\right]
{\rm d}y{\rm d}x\nonumber\\
&&\simeq-\frac {2\N}{\pi^2}\e^{-2\N}
\left(\int_{-\N}^\infty\e^{-x}\log \left[\B(x)\bB(x)\right]{\rm d}x\right)
\left(\int_{-\N}^\infty\e^{-y}\log \left[\tB(y)\tbB(y)\right]{\rm d}y\right).
\label{doublint}
\eea
Amazingly, the integrals in the final version are those we calculated already
in leading order
\begin{equation}
\left(\int_{-\N}^\infty\e^{-x}\log[\B\bB]dx\right)_\infty
=\frac 14\left(\frac{\pi^2}{3}+\frac 43 \phi^2\right).
\end{equation}
Hence
\bea
\int_{-\N}^\infty&\e^{-x}&\log[\B\bB]dx\nonumber\\
&\simeq&\left(\int_{-\N}^\infty\e^{-x}\log[\B\bB]dx\right)_\infty
-\frac 12\int_{-\N}^\infty\left[\ps'\log\B+\bps'\log\bB\right]dx,\nonumber\\
&\simeq&\left(\int_{-\N}^\infty\e^{-x}\log[\B\bB]dx\right)_\infty\nonumber\\
&&+\frac {\N}{\pi^2}\e^{-2\N}
\left(\int_{-\N}^\infty\e^{-x}\log \left[\B(x)\bB(x)\right]{\rm d}x\right)
\left(\int_{-\N}^\infty\e^{-y}\log \left[\tB(y)\tbB(y)\right]{\rm d}y\right).
\label{recursiveInt}
\eea
The final result reads
\begin{equation}
\int_{-\N}^\infty\e^{-x}\log[\B\bB]dx
\simeq\frac 14\left(\frac{\pi^2}{3}+\frac 43 \phi^2\right)
+\frac {\N}{16\pi^2}\e^{-2\N}\left(\frac{\pi^2}{3}+\frac 43 \phi^2\right)^2
\end{equation}
implying logarithmic corrections in $T$ for the free energy and the 
zero-field susceptibility
\bea
f&\simeq& e_0-\frac{T^2}{2\pi v}\left(
\left(\frac{\pi^2}{3}+\frac 43 \phi^2\right)
+\frac {\N}{4\pi^2}\e^{-2\N}\left(\frac{\pi^2}{3}+\frac 43 \phi^2\right)^2\right)
\quad\N=\ln(v/T),\quad (h<<T),\nonumber\\
&\simeq&e_0-\frac{\pi}{6v}T^2\left(1+\left(2\phi/\pi\right)^2\right)
\left(1+\frac {T^2\ln(v/T)}{12 v^2}\left(1+\left(2\phi/\pi\right)^2\right)\right),\\
\chi&=&\frac{3}{4\pi v}\left(1+\frac{2T^2\ln(v/T)}{3^4 J^2}\right).
\label{lowT}
\eea
This result is in complete agreement with the effective Hamiltonian analysis
of \cite{Lukyanov1997,SirkerBortz06} where also higher order terms were 
calculated. In this approach, the occurence of the logarithmic correction 
stems from the interference of the first-order contribution of the $T\bar T$ 
perturbation to the Gaussian Hamiltonian with the second-order contribution of 
the cosine of the bosonic fields.
The main motivation of our analysis is not the systematic extension of the
low temperature corrections. We intend to compute the entire low-lying 
excitation spectrum of the quantum transfer matrix and by the 
$\beta\leftrightarrow L$ correspondence the spectrum of the Hamiltonian up 
to order $1/L$ and $(\ln L)/L^3$.

\section{Critical indices and low-lying eigenvalues of the Heisenberg chain}

A major area of interest in statistical physics is the study of critical
properties of models at phase transitions and, more specifically, the
determination of the scaling dimensions $x$, from which the critical exponents
can be deduced. For the critical two-dimensional six-vertex model and the
associated spin-1/2 $XXZ$ quantum chain a wealth of information has been
collected by combinations of finite size studies of the eigenvalue spectrum
and scaling relations by conformal field theory.

The energy levels of the corresponding quantum model on a finite chain with
periodic boundary conditions scale with the system size $L$ like
(see \cite{Card88} and references therein)
\be
E_x - E_0 = \frac{2\pi}L vx + o\left(\frac 1L\right), \qquad
E_0 = Le_0 -\frac{\pi v}{6L}c,
\label{finiteSizeE}
\ee
where $E_0$ is the ground state energy, $E_x$ are the energies of the
low-lying excited states and $v$ is the sound velocity of the elementary
energy-momentum excitations. Equivalent formulae can be set up for the
spectrum of the row-to-row transfer matrix of the two-dimensional model.

For the critical six-vertex model the scaling dimensions have been treated by
analytical as well as numerical methods based on (\ref{finiteSizeE}). The
dimensions $x$ are given by the Gaussian
spectrum \cite{BogoIzKo86,BogoIzRe87,BogoIzKoBook}. The analytic method for
calculating finite-size corrections based on linear integral equations for
density functions of Bethe ansatz rapidities was applied to many models.
However, this method has the shortcoming that it fails for the higher spin-$S$
$XXZ$ chains and similar problems. There is an alternative procedure based on
non-linear integral equations (NLIE) for suitably chosen auxiliary functions
which proved to be applicable also in the hitherto problematic cases of higher
spin-$S$ chains, i.e.~for systems supporting filled seas of string
configurations. At the same time the NLIE method proved flexible
enough \cite{KWZ93} to give a comprehensive analytical study of the six-vertex
model spectrum.

The eigenvalues are obtained from non-linear integral equations \cite{KlumTH}
for auxiliary functions $\a$, and $\A=1+\a$
\begin{equation}
\log \frak a(x) 
= L\ln\tanh\frac x2 + C
+\int_{-\infty}^\infty\left[k(x-y)\log \frak A(y)
-k(x-y-i\pi+i\epsilon)\log\overline {\frak A}(y)\right]{\rm d}y,
\label{IntEqxxxHam}
\end{equation}
and the corresponding equation for $\ba$, and $\bA=1+\ba$ related by complex
conjugation. These equations are very similar to those in
(\ref{IntEqxxx}). The only difference is the change in the driving term
\begin{equation}
-\frac{v\beta}{\cosh x} \leftrightarrow L\ln\tanh\frac x2.
\end{equation}
Note that the asymptotics of both functions are the same exponentials with
different prefactors: $-2 v\beta \exp(-x)$ and $-2 L \exp(-x)$. The constant $C$
in (\ref{IntEqxxxHam}) is $i\pi\varphi /(\pi-\gamma)$ 
where $\varphi$ is the
twist angle of the generalized periodic boundary conditions. (For certain
excited states, the constant $C$ may take different values.)

The energy corresponding to a solution to the NLIE (\ref{IntEqxxxHam})
satisfies an integral expression very similar to (\ref{eigenxxx}). Therefore
the analysis of the finite size Hamiltonian can be carried out very closely to
the analysis of the low-temperature analysis of the previous section.

The main results for the general $XXZ$ chain were obtained already
in \cite{KWZ93} and may be summarized for the $1/L$ contributions to the 
energy and momentum eigenvalues as
\bea
\frac L{2\pi v}(E-L e_0)=-\frac{1}{2\pi^2}\left(
\int_{-\N}^\infty\e^{-x}\log\left[\B\bB(x)\right]dx
+\int_{-\N}^\infty\e^{-x}\log\left[\tB\tbB(x)\right]dx\right),
\nonumber\\
\frac L{2\pi}P=-\frac{1}{2\pi^2}\left(
\int_{-\N}^\infty\e^{-x}\log\left[\B\bB(x)\right]dx
-\int_{-\N}^\infty\e^{-x}\log\left[\tB\tbB(x)\right]dx\right).
\label{finitEnerg}
\eea
These formulae have a rather universal algebraic form as they are valid for
all eigenvalues of the Hamiltonian with suitably chosen contours! 

In \cite{KWZ93} the leading contribution to the integrals was calculated 
\begin{equation}
\int_{-\N}^\infty\e^{-x}\log\left[\Bpm\bBpm(x)\right]dx
=-{2\pi^2}\left(\frac{x(\varphi)\pm s(\varphi)}2 -\frac 1{24}\right) 
\end{equation}
where 
\bea
x(\varphi)&=&x_{S,m}(\varphi)=\frac{1-\gamma/\pi}2 S^2+\frac
1{2(1-\gamma/\pi)}\left(m-\frac\varphi\pi\right)^2\nonumber\\
s(\varphi)&=&S\left(m-\frac\varphi\pi\right)
\label{scalingdimspin}
\eea
are the `scaling dimension' and `conformal spin', and $S$ and $m$ are 
integers. From these results with $S=m=0$ and (\ref{finiteSizeE}) also the 
(effective) central charge follows
\begin{equation}
c(\varphi)=1-\frac{6(\varphi/\pi)^2}{1-\gamma/\pi}
\label{effcharge}
\end{equation}

For the special case of interest ($\Delta=1/2$) we are able to calculate the
next leading contribution to the integrals on the right hand side of
(\ref{finitEnerg}). We note that (\ref{recursiveInt}) holds with the result
\begin{equation}
\int_{-\N}^\infty\e^{-x}\log\left[\Bpm\bBpm(x)\right]dx
=\pi^2\left[1/{12}-(x(\varphi)\pm s(\varphi)) \right]
+\pi^2{\N}\e^{-2\N}\left[(1/{12}-x(\varphi))^2-s^2(\varphi))\right],
\end{equation}
where now
\begin{equation}
\N=\ln L \Rightarrow {\N}\e^{-2\N}=\frac{\ln L}{L^2}.
\end{equation}
These relations imply that the next-leading corrections
over the ($1/L$) conformal field theory results for the energy levels are
\begin{equation}
E_x-E_x^{CFT}=-2\pi v\left[(1/{12}-x(\varphi))^2-s^2(\varphi))\right]\frac{\ln L}{L^3},
\end{equation}
where $v$ is the velocity of the elementary excitations. Our results agree
with those of \cite{Lukyanov1997} and (slightly) generalize them to arbitrary
twisted boundary conditions.

\section{The $sl(2|1)$-invariant network model: Temperley-Lieb representation
with $\Delta=1/2$}

Here we investigate the critical properties of the statistical mechanical
network model derived in \cite{ChalkerCodd,Gruzberg} for the spin quantum Hall
transition which may occur in disordered superconductors with unbroken $SU(2)$
spin-rotation symmetry but broken time-reversal symmetry \cite{Altland}. Using
supersymmetry, and averaging over the random $SU(2)$ matrices a classical
3-state model on the square lattice was derived with local Boltzmann weights
resp. the $R$-matrix given as a superposition of the identity and the
projector onto the $sl(2|1)$-singlet state of nearest-neighbours.

Thus, the study of the localization-delocalization transition was mapped onto
the two-dimensional percolation problem. Furthermore, the local Boltzmann
weights derived in \cite{Gruzberg} are a realization of the Temperley-Lieb
algebra \cite{TemperleyLieb71,Baxt82b,PMBook91} and hence the model is related
to some (inhomogeneous) six-vertex model. For the symmetric case, the network
model in \cite{Gruzberg} is related to an integrable homogeneous six-vertex
model. 
In our representation theoretical treatment we give the identification of the
type of boundary conditions to be used for the six-vertex model which is
essential for the determination of the scaling dimensions in a finite-size
analysis of eigenvalues of transfer matrices and/or Hamiltonians.

The Temperley-Lieb algebra $TL(\lambda)$ is the unital associative algebra
over $\mathbb{C}$ generated by $e_1,\,e_2,\,\dots, e_{L-1}$ with relations
\begin{equation}
\begin{split}
  e_i^2  = &\; \lambda \; e_i, \\ 
 e_i \; e_{j} \; e_i   = & \;  e_i, \;\;\; \mbox{if}\ i, j\ \mbox{nearest 
neighbours}\\
 e_i\;e_j = & \;e_j\;e_i, \;\;\; \mbox{if}\ i, j\ \mbox{not nearest neighbours}
\label{TL-relations}
\end{split}
\end{equation} 
depending on the complex parameter $\lambda=2\Delta$, where $\Delta$ is the
anisotropy parameter of the anisotropic Heisenberg chain which is the standard
(faithful) representation of the Temperley-Lieb algebra.  The periodic closure
of $TL(\lambda)$ has one additional generator $e_L$ and relations
(\ref{TL-relations}) with 1 and $L$ beeing considered as neighbouring
indices. For a full description of the periodic (and more generally twisted)
closure we refer the reader to \cite{AufgebauerK10} and references therein.

According to the realization of $TL(\lambda)$ as a diagram algebra we use the
following graphical notation for the operators $e_i$:
\begin{equation}\mbox{\parbox{0.5cm}{$ e_i  $}
\hfill $=$ \hfill\parbox{7.5cm}{$\underbrace{~\epsfig{file=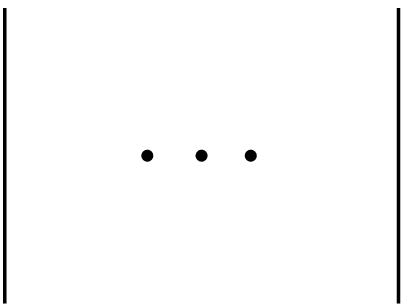,
      width= 2 cm}~}_{id^{\otimes
      (i-1)}}~\quad\epsfig{file=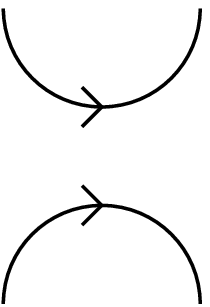, width= 1 cm}~ \;\quad
  \underbrace{~\epsfig{file=2nid.eps, width= 2 cm}~}_{id^{\otimes
     (L-i-1)}} $} }\label{bgra}
\end{equation}
where the two curved links correspond to the bra- and ket-states of the local
$sl(2|1)$ singlet.

Obviously, the parameter $\lambda$ appearing in (\ref{TL-relations}) has the
counterpart of a closed loop which for the model in \cite{Gruzberg} is 1
($=1-1+1$ due to the alternating standard and non-standard anti-commutation
relations for fermions on odd and even lattice sites).
These constructions may be generalized to the case of a $sl(n+1|n)$-invariant
network model, again of Temperley-Lieb type with $\Delta=1/2$, but now with a
total number of states per site $d=2n+1$. 

\begin{figure}[!htb]
  \begin{center}
    \includegraphics[width=0.7\textwidth]{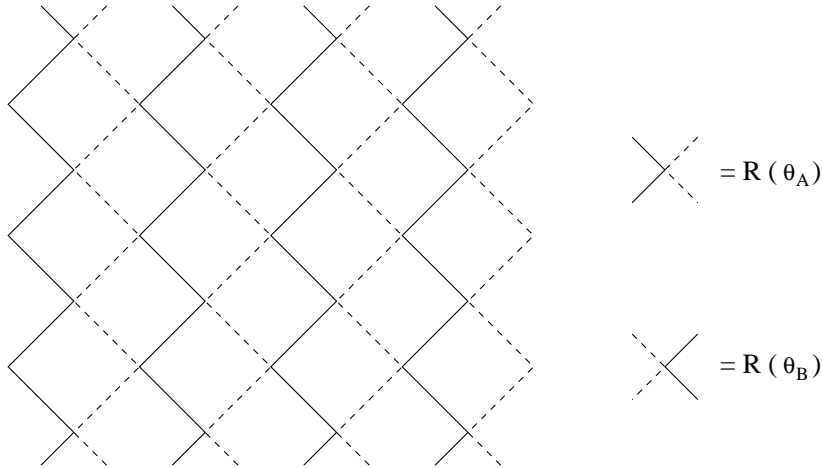}
    \caption{Depiction of the Chalker-Coddington model. Each vertical zig-zag
    line corresponds to a local quantum space. Solid and dashed lines
    correspond to dual representations of $sl(2|1)$. In the general,
    inhomogeneous case, there are independent local interactions $R(\Theta_A)$
    and $R(\Theta_B)$ on even and odd rows.}
    \label{fig:Gruz}
  \end{center}
\end{figure}

\begin{figure}[!htb]
  \begin{center} \includegraphics[width=0.7\textwidth]{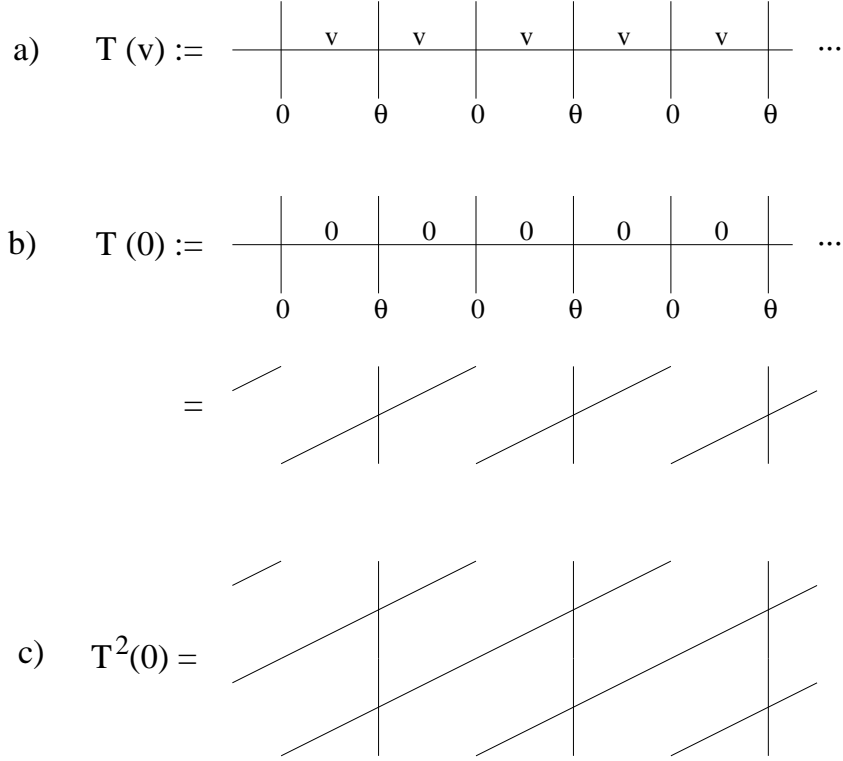} \caption{a)
 Graphical depiction of a row-to-row transfer matrix $T(v)$ with spectral
 parameter $v$ associated with the horizontal line and alternatingly 0 and
 $\Theta$ associated with the vertical lines. Interactions are formulated for
 each vertex with local Boltzmann weight given by an $R$-matrix $R(u)$ where
 $u$ is taken as the difference of the two spectral parameters of the
 intersecting lines. This construction renders the transfer matrix
 integrable. Note the factorization taking place for $v=0$ at every second
 vertex as $R(0)$ reduces to the identity, a) and b).}
\label{fig:TransferM} \end{center}
\end{figure}

The statistical mechanical model of \cite{Gruzberg} has a staggered structure
with two types of local Boltzmann weights alternating from row to row,
cf. Fig.~\ref{fig:Gruz}. The local Boltzmann weights can be given in terms
of so-called $R$-matrices which are defined by
\begin{equation}
R_{j,j+1}(u):=\sin(\gamma-u)+\sin(u) e_j
\label{Rmatrix}
\end{equation}
where $\gamma=\pi/3$ and $u$ takes certain values $\Theta_A$ and $\Theta_B$
alternatingly from row to row.
For general parameters $\Theta_{A,B}$ the model is not exactly
solvable. However, for identical interaction parameters
$\Theta:=\Theta_A=\Theta_B$, the row-to-row transfer matrices belong to a
commuting family of transfer matrices defined on a square lattice with the
same width and height as in Fig.~\ref{fig:Gruz} but with horizontal and
vertical lines to which spectral parameters are associated. On the vertical
lines we place alternatingly 0 and $\Theta$. Now consider a single row with
spectral parameter $v$ attached to the horizontal line. The corresponding
transfer matrix is denoted by $T(v)$ and consists of an alternating product of
$R(v-0)$ and $R(v-\Theta)$ matrices, see Fig.~\ref{fig:TransferM} a). All
matrices $T(v)$ with arbitrary complex argument $v$ commute due to the
Yang-Baxter relation satisfied by (\ref{Rmatrix}). For $v=0$ the lattice in
Fig.~\ref{fig:Gruz} is recovered. This is most easily understood by observing
that $R(0)$ is proportional to the identity operator and $T(0)$ factors as
shown in Fig.~\ref{fig:TransferM} b). The product of two transfer matrices
$T(0)$, cf. Fig.~\ref{fig:TransferM} b), corresponds to two successive rows in
Fig.~\ref{fig:Gruz} times a translation by two lattice units. For a lattice of
square shape the translation operators reduce to the identity and the
partition function of Fig.~\ref{fig:Gruz} is identical to
$\textnormal{Tr}\big[T(0)^L\big]$.

Instead of working out the critical properties directly for the model depicted
in Fig.~\ref{fig:Gruz} (with $\Theta=\Theta_A=\Theta_B$ we use the
integrability of the transfer matrix $T(v)$. Due to the commutativity, the
eigenstates of $T(v)$ are independent of $v$. The largest eigenvalue of $T(0)$
is realized by the eigenstate with the lowest eigenvalue of the corresponding
Hamiltonian $H$. Note that also the product $T_2(v):=T(v)T(v+\Theta)$ is a
commuting set of matrices generated by the complex variable $v$. For $v=0$ and
by use of `unitarity' for the $R$-matrix, this matrix can be shown to reduce
to the translation operator by two lattice sites. The logarithmic derivative
of $T_2(v)$ is nothing but
\begin{equation}
H=-\sum_{i=1}^L e_i
\end{equation}
The critical properties of the system, e.g. correlation functions of local
operators, are described by the scaling dimensions which are obtained from the
finite-size data of the low-lying excitations of this Hamiltonian with
periodic boundary conditions or with twist angle $\pi$. 
In the dense loop representation of the model, periodic boundary conditions
render closed loops to evaluate to 1 (= number of bosonic states per site -
number of fermionic states per site), for $\pi$-twisted boundary condition
such loops evaluate to the total number of states per site $d$.

As shown in \cite{AufgebauerK10} the sub-representations of the above
Temperley-Lieb model appear as sub-representations in the standard reference
Heisenberg chain, however with a new twist angle $\varphi$ appearing. The
representations are constructed by attaching $k$-many singlets ($k \le L/2$)
to a pseudo-vacuum on a chain of length $N=L-2k$.  The dimension of the space
$\Omega_N^p$ of pseudo-vacua of length $N$ for periodic boundary conditions is
given by (see appendix)
\begin{equation}
\mbox{dim}(\Omega_N^p)(d)=\left ( \frac{d}{2} +\frac{ \sqrt{d^2-4}}{2}
\right )^N + \left ( \frac{d}{2} - \frac{\sqrt{d^2-4}}{2} \right
)^N,\label{dimension ref}
\end{equation}
which specializes for a 3-dimensional local Hilbert space to
$\mbox{dim}(\Omega_N^p)(3)=\left[(3+\sqrt 5)/2\right]^N+\left[(3-\sqrt 5)/2\right]^N$.

In this space the translation operator shifting the system by two lattice
sites can be diagonalized yielding momentum eigenvalues $P=l\cdot 2\pi/S$
where $l=0,...,S-1$ ($N=2S$). The energy eigenvalues of the Temperley-Lieb
model in the sub-representation with $k$-many singlets and momentum $l\cdot
2\pi/S$ reference state are identical to those of the $XXZ$ quantum chain with
$k$-many flipped spins (i.e. $L-k$ up and $k$ down spins with magnetization
$S=N/2$) and twist angle
\begin{equation}
\varphi=\frac P2=l\cdot\frac\pi{S}.
\label{twistnonzero}
\end{equation}
Note that the dependence of the eigenvalues of the $XXZ$ reference system on
$\varphi$ is $\pi$-periodic.  The quantum number $l$ takes integer values for
periodic and $\pi$-twisted boundary conditions of the Temperley-Lieb model,
except for the $\pi$-twist and an odd number of fermions in the
pseudo-vacuum. In such a case, the set of allowed values for $l$ is shifted by
$1/2$ ($l=1/2, 3/2,..., S-1/2$) which will not be pursued furthermore.

The case $S=0$ ($k=L/2$, $N=0$) is special. Here the relation between the
twist angles is given by equating macroscopic loops
\begin{equation}
1\mp 1+1 = \e^{i\varphi}+\e^{-i\varphi},
\label{macroscoploop}
\end{equation}
for periodic boundary conditions and $\pi$-twist. Explicitly we find 
\begin{equation}
\varphi=\pi/3 \quad\hbox{resp.}\quad i\ln\left((3+\sqrt 5)/2\right),
\label{twistzero}
\end{equation}
a real/imaginary value for periodic boundary conditions and $\pi$-twist.  
(For the general $sl(n+1|n)$ case, the left hand side in (\ref{macroscoploop})
is to be replaced by 1 or $d$ for periodic or $\pi$-twisted boundary
conditions.)

We read off the central charge from (\ref{effcharge}) with $\gamma=\pi/3$ and
(\ref{twistzero}) as $c=0$ for periodic boundary conditions and
$c=1+\left(\frac 3\pi\ln\left[(3+\sqrt 5)/2\right]\right)^2$ for the
$\pi$-twist. The scaling dimensions corresponding to low-lying excitations
in the sub-representation with $k$-many singlets appended to a length $N$
pseudo-vacuum ($N=L-2k$) are obtained from (\ref{scalingdimspin})
\bea
x&=&\frac L{2\pi v}\left[E_L(\varphi)-E_{L,0}(\pi/3)\right]
=\frac L{2\pi v}\left(\left[E_L(\varphi)-E_{L,0}(0)\right]-
\left[E_L(\pi/3)-E_{L,0}(0)\right]\right)\\
&=&x_{S,m}(\varphi)-x_{0,0}(\pi/3)=
\frac{4\,S^2+9(m-\varphi/\pi)^2-1}{12},
\label{scalrelTL}
\eea
where the twist angle $\varphi$ is as in (\ref{twistnonzero}) and the true
ground-state corresponds to $S=m=0$ and twist angle in (\ref{twistzero}).

Next we give the multiplicities of the momentum values in the space
$\Omega_N^p$ of all pseudo-vacua of length $N$ for periodic boundary
conditions.  First, the dimension of this space is (see appendix)
\begin{equation}
d(N):=\mbox{dim}(\Omega_N^p)=\left ( \frac{d}{2} +\frac{ \sqrt{d^2-4}}{2}
\right )^N + \left ( \frac{d}{2} - \frac{\sqrt{d^2-4}}{2} \right
)^N. 
\end{equation}
The subspace of $\Omega_N^p$ of states with an odd number of fermions has dimension
(see appendix)
\begin{equation}
d^o(N)=\frac{d(N)}2-\cos\left(\frac\pi 3 N\right).
\label{oddferm}
\end{equation}
The eigenstates of the translation operator by two sites in $\Omega_N^p$ are
constructed from the orbits with periods $p$ (which are divisors of $N$).
Here we follow closely the treatment in \cite{AufgebauerK10}, the main
difference is that we now deal with a translation operator by two lattice
sites and hence an orbit with period $p$ covers $2p$ sites.  The dimensions of
the orbits are denoted by $\sigma(p)$. Similary, we denote by $\sigma^o(p)$
the dimension of the subspace of the orbit with period $p$ containing states
with an odd number of fermions. The dimensions of the orbits are related to
the dimension of all pseudo-vacua by
\begin{equation}
\sum_{p\,|\,S}\sigma(p)=d(2S),\label{sump}
\end{equation} 
as well as
\begin{equation}
\sum_{p\,|\,M}\sigma^o(2^mp)=d^o(2^{m+1}M),\label{sump0}
\end{equation}
where $m$ and (odd) $M$ are uniquely defined integers such that $S=2^mM$.

These equations are uniquely solved by
\begin{equation}
{\sigma}(S)=\sum_{r\,|\,S}\mu\left(\frac{S}{r}\right)\;d(2r),
\label{sigma2}
\end{equation}
and 
\begin{equation}
{\sigma}^o(2^mM)=\sum_{r\,|\,M}\mu\left(\frac{M}{r}\right)\;d^o(2^{m+1}r)
\end{equation}
where $\mu$ is the M\"obius function defined for integers $q$ by
\begin{equation*}
\mu(q):=\begin{cases}
1,&\text{for $q=1$},\\
(-1)^s,&\text{if $q$ is the product of $s$ distinct primes},\\
0,&\text{else}.\end{cases} 
\end{equation*}
For periodic boundary conditions the multiplicity of the momentum
$\frac{2\pi}{S}l$ in $\Omega_N^p$ is
\begin{equation}
 M\left (\frac{2\pi}{S}l,\Omega^p_N\right )=\sum\limits_{r
 |{\rm gcd}(S,l)}\frac{r}{S}\left[{\sigma}\left(\frac Sr\right)
-\frac{1+(-1)^r}2{\sigma}^o\left(\frac Sr\right)\right]
+\sum\limits_{\substack{r|{\rm gcd}(S,2l)\\r\, {\rm even},\ 2l/r\, {\rm odd}}}\frac{r}{S}\,\sigma^o\left(\frac Sr\right).
\label{multiplic}
\end{equation}

There are some interesting applications of the above results. 
The decay of
2-point functions at the critical point are governed by the lowest values of
$x$ with the right quantum numbers $S, m, l$, where, however, the application
of selection rules is obscured as the space of pseudo-vacua $\Omega_N^p$ is
rather high dimensional and `violates' simple conservation numbers.

With respect to the localization-delocalization exponent $\nu$ the result
$4/3$ was found in \cite{Gruzberg} from the well established general RG
scaling relation
\begin{equation}
\nu=\frac 1{2-x},
\end{equation}
where the `relevant' scaling dimension was identified as that of the 2-hull
operator which corresponds to $S=2$ and $m=l=0$ in (\ref{scalrelTL}). (For $l=1$ the value of $\nu$ is $1.\overline 7$.) 

We like to note that the energy levels of the model by \cite{Gade} show
stronger logarithmic corrections, $1/(L\ln L)$, \cite{EssFrahmSaleur} than
observed here, $(\ln L)/L^3$, cf.~Sect.~3.
This indicates that the two models correspond to rather different
universality classes.

\begin{figure}[!htb]
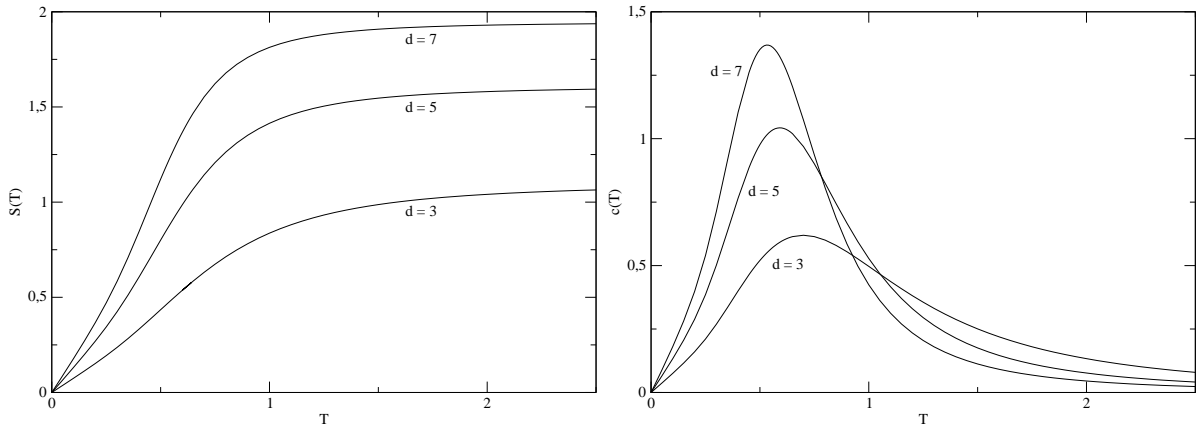

  \begin{center}
    \includegraphics*[width=0.49\textwidth]{Entrop_D_5.eps}
    \includegraphics*[width=0.49\textwidth]{Spezif_D_5.eps}
    \caption{(a) Entropy $S(T)$ and (b) specific heat $c(T)$ data versus
    temperature $T$ for the Temperley-Lieb Hamiltonians with $\Delta=1/2$ 
    corresponding to the $sl(n+1|n)$ invariant network models with $n=1,
    2, 3$ $(d=3, 5, 7)$.}  
\label{fig:specsusy} 
\end{center}
\end{figure}

Finally, we recall from \cite{AufgebauerK10} the finite temperature relation
of a Temperley-Lieb vertex model with local Hilbert space of dimension
$d=2n+1$ to the Heisenberg chain at same temperature {\it plus an effective
magnetic field} $\tilde h$ proportional to the temperature
\begin{equation}
f(T)=f_{XXZ}(T,\tilde h),\quad 2\cosh\frac{\beta \tilde h}2=d.
\label{freeenergy}
\end{equation}
In Fig.~\ref{fig:specsusy} we show results for the entropy and the specific
heat of the first three cases of the Hamiltonians for $n=1, 2, 3$. Note that
the thermodynamical data at low-temperatures are very similar for all cases
with $c(T)$ exhibiting a proportionality to temperature with slope $\pi c/3v$
with central charge $c$ given by (\ref{effcharge}) and imaginary angle
$\varphi$ with $\cos\varphi=d/2$.  At higher temperatures the curves deviate
from each other. This is expected on physical grounds as in the high
temperature regime the entropy has to approach $\ln d=\ln(2n+1)$.

If we impose $\pi$-twisted boundary conditions in imaginary time direction,
i.e.~if we define a pseudo-partition function by use of the super-trace, we
find a vanishing free energy for all temperatures consistent with the
low-temperature asymptotics with central charge $c=0$.

\section{Conclusion}
In this paper we derived the complete low-lying energy spectrum of the quantum
spin chain related to the $sl(n+1|n)$ supersymmetric network
of \cite{Gruzberg}. Our analysis consists of two major steps.

First, we established a proper Temperley-Lieb equivalence of the quantum spin
chain with periodic/twisted boundary conditions to a spin-1/2 $XXZ$ chain with
anisotropy $\Delta=1/2$ and suitable boundary conditions.
We presented a representation theoretical approach as a simple alternative to the evaluation of the scaling dimensions and multiplicities in \cite{ReadSaleur01,RichardJacobsen}. Second, the leading logarithmic
corrections to the conformal spectrum were derived.
Our results indicate that the models of \cite{Gruzberg} and \cite{Gade} belong
to different universality classes.

The calculations are based on an asymptotic analysis of non-linear integral
equations for the finite temperature and finite size properties of the
Heisenberg spin chain. These equations are exact for arbitrary size.  By these
means, also the thermodynamical properties of the supersymmetric Hamiltonians
of Temperley-Lieb type were computed for arbitrary temperature. The specific
heat exhibits a linear low-temperature behaviour with a slope corresponding
to the central charge in the $\pi$-twisted sector.

\section*{Acknowledgments}
The authors acknowledge valuable discussions with H. Boos, F. Essler,
H. Frahm, F. G\"ohmann, J.L. Jacobsen and C. Trippe.  B.A., M.B. and
W.N. acknowledge financial support by the DFG research training group 1052 and
by the VolkswagenStiftung.

\section*{Appendix}

In this appendix we give a short account of the derivation of the multiplicity
formulas and a proof of the completeness of states.
All formulae are to be understood as for the generic Temperley-Lieb representations in the limit of $\lambda=2 \Delta \to 1$ avoiding the explicit analysis of reducible but indecomposable representations. 

We begin with the characterization of the space of pseudo-vacua $\Omega_N^p$
containing all states $\omega$ on a chain of length $N$ with periodic boundary
conditions annihilated by all operators $b_1$,...,$b_N$. If we label the
canonical basis states of the local spaces $V$ with $j=1,...,d\ (=2n+1)$ and
define $U:V\to V$ by $U_{i,j}:=\delta_{i,d+1-j}$, the space $\Omega_N^p$ is
mapped by $U\otimes{\rm id}\otimes U\otimes{\rm id}\otimes ...$ onto a space
of states that are annihilated by the application of dual states
$\left<\psi\right|:=\sum_{j=1}^d\left<j,d+1-j\right|$ and
$\left<\tilde\psi\right|:=\sum_{j=1}^d(-1)^{j+1}\left<j,d+1-j\right|$ on
(odd,even)- and (even,odd)-indexed nearest-neighbours, respectively.

We introduce a $q$-deformed version of this space by demanding annihilation by
$\sum_{j=1}^dq^{j-(d+1)/2}\left<j,d+1-j\right|$ and
$\sum_{j=1}^d(-1)^{j+1}q^{j-(d+1)/2}\left<j,d+1-j\right|$.  The dimension of
the space is independent of $q$ (see \cite{AufgebauerK10}) and can be easily
enumerated in the limit $q\to\infty$. The condition for the states is simply
that the sequence $(d,1)$ never occurs. The counting problem is now reduced to
the evaluation of traces of powers of the $d\times d$ adjacency matrix $A$
with entries 1 everywhere except for one 0 at the position $(d,1)$. The only
non-zero eigenvalues are $\alpha_\pm=(d\pm\sqrt{d^2-4})/2$, where
$\alpha_-=1/\alpha_+$. Hence
\begin{equation}
d(N):={\rm Tr} A^N=\alpha_+^N+\alpha_-^N,
\end{equation}
which proves (\ref{dimension ref}) for $N>0$.

Defining the diagonal $d\times d$ matrix $R:={\rm diag}(+1,-1,+1,-1,..)$ we
find
\begin{equation}
{\rm Tr} (AR)^N=d^e(N)-d^o(N),
\end{equation}
where $d^e(N)$ and $d^o(N)$ are the dimensions of the subspaces with even and
odd numbers of fermions, respectively. The only non-zero eigenvalues of $AR$
are $(1\pm{\rm i}\sqrt 3)/2$, hence $d^e(N)-d^o(N)=2\cos\left(\frac\pi
3N\right)$ proving (\ref{oddferm}).

Finally, we like to count the number of states covered by the constructed
representations. The dimension of each $k$-sector ($k=0,1,...,L/2$) is
$L\choose k$ and appears $d(L-2k)$ times for $k<L/2$ and once for
$k=L/2$. Hence the number of states is
\begin{equation}
\sum_{k=0}^{L/2-1} {L\choose k} \left(\alpha_+^{L-2k}+\alpha_+^{-(L-2k)}\right)
+{L\choose L/2}=\sum_{k=0}^{L} {L\choose k}\alpha_+^{L-2k}
=\alpha_+^L(1+\alpha_+^{-2})^L=d^L.
\end{equation}

\end{document}